\documentclass[aps,prd,10pt,twocolumn]{revtex4}

\usepackage{amsmath,amssymb,amsfonts,dcolumn,color,graphicx,graphics,latexsym,placeins,epsfig}
\usepackage{epsfig}
\usepackage{bm}
\usepackage{latexsym}
\usepackage{natbib}
\usepackage{url}
\usepackage{dcolumn}
\usepackage{color}
\usepackage{amsfonts,amssymb,amsmath}
\usepackage{graphicx,epsfig}
\usepackage{psfrag}
\usepackage{subfigure}
\usepackage{hyperref}
\hypersetup{colorlinks=true}

\newcommand{\be}{\begin{equation}}
\newcommand{\ee}{\end{equation}}
\newcommand{\ba}{\begin{eqnarray}}
\newcommand{\ea}{\end{eqnarray}}

\newcommand{\p}{\prime}

\begin{document}

\title{Constraints on Born-Infeld gravity from the speed of gravitational waves after GW170817 and GRB 170817A}

\author{Soumya Jana\footnote{sjana@prl.res.in}}
\author{Girish Kumar Chakravarty\footnote{girish20@prl.res.in}}
\author{Subhendra Mohanty\footnote{mohanty@prl.res.in}}
\affiliation{\rm Theoretical Physics Division, Physical Research Laboratory, Ahmedabad 380009, India }


\begin{abstract}
\noindent  The observations of gravitational waves from the binary neutron star merger event GW170817 and the subsequent observation of its electromagnetic counterparts from the gamma-ray burst GRB 170817A provide us a significant opportunity to study theories of gravity beyond general relativity.  An important outcome of these observations is that they constrain the difference between the speed of gravity and the speed of light to less than $10^{-15}c$. Also, the time delay between the arrivals of gravitational waves at different detectors constrains the speed of gravity at the Earth to be in the range $0.55c < v_{gw} < 1.42c$. We use these results to constrain a widely studied modified theory of gravity: Eddington-inspired Born-Infeld (EiBI) gravity. We show that, in EiBI theory, the speed of gravitational waves in matter deviates from $c$. From the time delay in arrival of gravitational wave signals at Earth-based detectors, we obtain the bound on the theory parameter $\kappa$ as $\vert\kappa\vert \lesssim 10^{21}\, m^2$. Similarly, from the time delay between the signals of GW170817 and GRB 170817A, in
a background Friedmann-Robertson-Walker universe, we obtain $\vert \kappa \vert \lesssim 10^{37}\, m^2$. Although the bounds on $\kappa$ are weak compared to other earlier bounds from the study of neutron stars, stellar evolution, primordial nucleosynthesis, etc., our bounds are from the direct observations and thus worth noting. 
\end{abstract}


\maketitle

\section{\bf Introduction} 


General relativity (GR) is extremely successful as a classical theory 
of gravity. Over the years, it has been scrutinized in vacuum or in the weak-field regime through several precision tests, and no significant deviation from GR has been found \cite{Will2014}. 
Still, there exist many unsolved puzzles in GR such as the problem of singularities (which is expected to be resolved by quantum gravity), understanding the dark matter and dark energy, etc.  
In order to address some of these problems, many researchers actively pursue 
modified gravity theories in the classical domain which deviate from GR inside matter distributions, or in the strong-field regime. One such modification is inspired by the well-known Born-Infeld electrodynamics \cite{born} where, even at the classical level, it is possible to avoid the infinity in the electric field at the location of a point charge. Deser and Gibbons \cite{desgib} first suggested a gravity theory in the metric formalism consisting of a similar determinantal structure $\sqrt{-|g_{\mu\nu}+\kappa R_{\mu\nu}|}$ as in the action of Born-Infeld electrodynamics. In fact, the determinantal form of the gravitational action is not a new concept;
it existed earlier in Eddington's reformulation of GR in de Sitter spacetime \cite{edd}. This is essentially an affine formalism where the affine connection is the basic variable instead of the metric; however, the coupling of matter to gravity remained a problem.

Later, the Palatini (metric-affine) formulation in Born-Infeld gravity was introduced by Vollick \cite{vollick}. He worked on various related aspects and also introduced a nontrivial and somewhat artificial way of coupling matter in such a theory \cite{vollick2,vollick3}. More recently, Ba\~{n}ados and Ferreira  \cite{banados} have come up with a formulation, popularly known as Eddington-inspired Born-Infeld (EiBI) gravity, where the matter coupling is different and simpler compared to Vollick's proposal. For a recent review on Born-Infeld gravity, see Ref. \cite{jimenez2017} and for its cosmological, astrophysical, and other applications see Refs. \cite{cardoso,delsate,pani,scargill,cho,chokim,escamilla,linear.perturbation,wei,jana,jana2,jana4,jana2017,rajibul,sotani,eibiwormhole,BTZ_typesoln,casanellas,avelino,sham,sham2,structure.exotic.star,sotani.neutron.star, sotani.stellar.oscillations,sotani.magnetic.star,wei,sotani,odintsov,fernandes,
latorre2017} and references therein.

The EiBI theory reduces to GR in vacuum but differs in the presence of matter. Therefore most stringent tests of this theory come from neutron stars and stellar evolution~\cite{cardoso,casanellas}. However, the recent direct detection of gravitational wave (GW) signals from the binary black hole and neutron star mergers \cite{abbott,gw170817} not only provide direct confirmation of one of the major predictions of GR, but also give a platform to probe the gravity deeper into the strong-field regime. The time delay between the observations of gravitational wave signals from the binary neutron star merger GW170817 and the observation of its associated electromagnetic counterparts from the gamma-ray burst GRB-170817A constrain the difference between the speed of gravity and the speed of light to less than $10^{-15}c$; more specifically, $|\frac{v_{gw}-c}{c}| \lesssim 10^{-15}$~\cite{abbot_2017oct}. Also, the observations of GWs at several Earth-based detectors constrain the speed of gravity between $0.55c < v_{gw} < 1.42c$~\cite{cornish2017}. These observations have been used to constrain the theories of gravity beyond GR~\cite{lombriser2016,lombriser2017,Zumalacarregui2016,shoemaker2017,baker1710,Sakstein:2017xjx,Creminelli:2017sry,Ezquiaga:2017ekz,1711.00492}. In this paper, we point out that another theory where the speed of gravitational waves deviates from the speed of light is EiBI gravity. Note that, like GR, also in EiBI gravity, the graviton is massless and there are only two polarization modes. Although, gravitational waves propagate in vacuum in exactly the same manner as in GR, there should be some differences due to matter distributions.



Various upper bounds on the theory parameter ($\kappa$, illustrated in the next section) of EiBI gravity exist in the literature from astrophysical and cosmological observations. For example, a strong constraint on the theory with $\kappa > 0$ and $\kappa \lesssim 10^9 $ m$^2$ comes from the existence of self-gravitating compact objects like neutron stars \cite{cardoso}. Stellar equilibrium and the evolution of the Sun lead to $\vert\kappa\vert \lesssim 2\times 10^{14}$ m$^2$ \cite{casanellas}. Assuming $\kappa>0$, the constraint $\kappa \lesssim 3.5 \times 10^{17}$ m$^2$ was obtained from the conditions for primordial nucleosynthesis 
\cite{avelino}. Requiring that the electromagnetic force dominates over the gravitational force at the subatomic scale leads to the very strong constraint $\vert \kappa \vert \lesssim 6\times 10^5$ m$^2$ \cite{nuclear.test}. Note that we express $\kappa$ in dimensions of [length]${}^2$, i.e., in units of $m^2$. In most of the literature, the units used are kg$^{-1}$m$^5$s$^{-2}$ with the assumption of $8\pi G=1$ and, therefore, one should divide the numbers by $8\pi G$ for the conversion. However, most of these bounds are somewhat indirect. In this article, we obtain constraints on $\kappa$ from the bounds on the speed of gravitational waves as mentioned above. In Sec. II, we obtain the gravitational-wave propagation equation in the background of Earth's gravitational field, and using this we put a constraint on $\kappa$ from the time delay in the arrival of gravitational-wave signals at widely separated earth based detectors. In Section III, we put constraint on $\kappa$ from the time delay between the gravitational wave signal from the recently detected neutron star merger event GW170817 and the electromagnetic signal from the associated $\gamma$-ray burst event GRB 170817A. Finally, we conclude and summarize our results in Sec. IV.

\section{Speed of gravity through the Earth in Born-Infeld gravity}
First, we briefly recall EiBI gravity. The central feature here is the existence of a physical metric which couples to matter, and another auxiliary metric which is not used for matter couplings. One needs to solve for both metrics through the field equations. The action for the theory developed in Ref. \cite{banados} is given as
\ba
S_{BI}(g,\Gamma, \Psi) &=&\frac{c^3}{8\pi G\kappa}\int d^4 x \left[\sqrt{-\vert g_{\mu\nu} +\kappa R_{\mu\nu}(\Gamma)\vert}\right.\nonumber\\
&& \left. -\lambda \sqrt{-g} \right]+ S_M (g, \Psi)
\ea
where $\lambda= \kappa \Lambda +1$, with $\Lambda$ being the cosmological constant. As mentioned earlier, $\kappa$ is the constant parameter of the theory with dimensions of [length]${}^2$ and, for sufficiently small $\kappa$, the action reduces to the known Einstein-Hilbert action. 
Variation with respect to $\Gamma$ (assuming symmetric $\Gamma^{\rho}_{\mu\nu}$ and $R_{\mu\nu}$) gives
\begin{equation}
q_{\mu \nu}=g_{\mu\nu} + \kappa R_{\mu\nu}(\Gamma),
\label{eq:gammavarn}
\end{equation}
where $q_{\mu\nu}$ is called the auxiliary metric which satisfies the compatibility condition 
$\nabla_{\alpha} (\sqrt{-q}q^{\mu\nu})=0$ with respect to $\Gamma$, which gives
\begin{equation}
  \Gamma^{\rho}_{\mu\nu}=\frac{1}{2}q^{\rho\alpha}\left(q_{\alpha\mu,\nu}+q_{\nu\alpha,\mu}-q_{\mu\nu,\alpha}\right).
  \label{eq:Gammadefn}
\end{equation}
Variation with respect to $g_{\mu\nu}$ gives the field equation
\begin{equation}
\sqrt{-q} q^{\mu\nu} = \lambda \sqrt{-g}g^{\mu\nu}-\frac{8\pi G}{c^4}\kappa \sqrt{-g} T^{\mu\nu},
\label{eq:gvarn}
\end{equation}
where the $T^{\mu\nu}$ components are in the coordinate frame.

 Now we derive the equation for the propagation of gravitational waves in the background of Earth's gravity. For the background field equations,
we work in the Newtonian limit or, more precisely, the nonrelativistic limit of EiBI theory \cite{banados}.
We consider a time-independent metric,
\be
ds^2=-\left(1+\frac{2\Phi}{c^2}\right)c^2dt^2+\left(1-\frac{2\Phi}{c^2}\right)\delta_{ij}dx^{i}dx^{j},
\label{eq:newton_line_element1}
\ee
coupled to the energy-momentum tensor $T^{\mu\nu}\approx \rho c^2u^{\mu}u^{\nu}$, in the comoving frame. Here $\rho$ is the matter density and $p$ is the pressure. In the nonrelativistic limit $p<<\rho c^2$ and $\Phi$ is the Newtonian potential. Also, here, the cosmological constant $\Lambda$ is irrelevant. We additionally assume a time-independent auxiliary metric,
\be
ds^2_q=-\left(1+\frac{2\Phi_q}{c^2}\right)c^2dt^2+\left(1-\frac{2\Phi_q}{c^2}\right)\delta_{ij}dx^{i}dx^{j}\,.
\label{eq:newton_line_element2}
\ee
From the $\mu=0,\nu=0$ (temporal indices) field equations [Eqs.~(\ref{eq:gvarn}) and (\ref{eq:gammavarn})], we have
\begin{eqnarray}
\Phi=\Phi_q+2\pi G \kappa\rho \label{eq:linearized_eibif1},\\
\Phi=\Phi_q+\frac{\kappa}{2}\nabla^2\Phi_q ,
\label{eq:linearized_eibif2}
\end{eqnarray}
where we keep only linear terms in $\Phi,\Phi_q$, and $\rho$. In Eq.~(\ref{eq:linearized_eibif2}), we used $R_{00}(q)=\nabla^2\Phi_q/c^2$. From Eqs.~(\ref{eq:linearized_eibif1}) and (\ref{eq:linearized_eibif2}) we get
\be
\nabla^2\Phi_q= 4\pi G \rho.
\label{eq:linearized_eibif3}
\ee
Taking the Laplacian of both sides of Eq.~(\ref{eq:linearized_eibif1}) we get
\be
\nabla^2 \Phi= \nabla^2 \Phi_q +2\pi G\kappa \nabla^2 \rho,
\label{eq:linearized_eibif4}  
\ee
and using Eq.~(\ref{eq:linearized_eibif3}) in Eq.~(\ref{eq:linearized_eibif4}), we finally get the modified Poisson equation
\be
\nabla^2\Phi=4\pi G \rho +2\pi G \kappa\nabla^2\rho.
\label{eq:modpoisson}
\ee
Using this modified Poisson equation, it was shown in Ref. \cite{cardoso} that this theory supports a stable pressureless neutron star. They it was also shown that nonrelativistic dust collapse does not lead to a singularity for $\kappa>0$, which is a completely different result from that in Newtonian gravity. The second term in the modified Poisson equation may play a role of repulsive role and may be important in a highly dense region of matter. However, for a nearly constant matter density (such as the Earth), this effect is negligible and
we get back Newtonian gravity. But we will see below that, even for the Earth, there will be a nonzero contribution in the equation of gravitational-wave propagation.

Now, in the presence of gravitational waves, the perturbed line elements will take the following forms:
%
\hspace{-0.5cm}\begin{eqnarray}
  ds^2 = -\left(1+\frac{2\Phi}{c^2}\right)c^2dt^2+\left(1-\frac{2\Phi}{c^2}\right)(\delta_{ij}+h_{ij})dx^{i}dx^{j},
  \label{eq:perturbed_physical}\nonumber\\
  ds^2_q=-\left(1+\frac{2\Phi_q}{c^2}\right)c^2dt^2+\left(1-\frac{2\Phi_q}{c^2}\right)(\delta_{ij}+\gamma_{ij})dx^{i}dx^{j},
  \label{eq:perturbed_auxiliary}\nonumber
\end{eqnarray}
where $h_{ij}(\vec{x},t)$ and $\gamma_{ij}(\vec{x},t)$ are transverse and traceless, i.e., $h_{ii}=\gamma_{ii}=0$ and $\partial_{i}h^{ij}=\partial_{i}\gamma^{ij}=0$. In our computations, we keep the terms in first order of $h_{ij}$ and $\gamma_{ij}$ as well as $\Phi$ and $\Phi_q$. However,
we keep the terms like ``$\Phi h_{ij}$" to capture the effect of Earth's gravity. The following identities are used to construct the perturbed
field equations:
\begin{eqnarray*}
  \sqrt{-g}&\approx& c(1-\frac{2\Phi}{c^2})+\mathcal{O}(h^2),\\
  \sqrt{-q}&\approx& c(1-\frac{2\Phi_q}{c^2})+\mathcal{O}(\gamma^2),\\
  g^{ij}&\approx& (1+\frac{2\Phi}{c^2})\left(\delta^{ij}-h^{ij}\right), \\
  q^{ij}&\approx& (1+\frac{2\Phi_q}{c^2})\left(\delta^{ij}-\gamma^{ij}\right).
\end{eqnarray*}
The $\mu=i$, $\nu=j$ (spatial indices) field equation from $g$ variation [Eq.~(\ref{eq:gvarn})] becomes
\ba
  &&\left(1-\frac{2\Phi_q}{c^2}\right)\left(1+\frac{2\Phi_q}{c^2}\right)\left(\delta^{ij}-\gamma^{ij}\right)\nonumber\\
  && \hspace{2.5cm} =\left(1-\frac{2\Phi}{c^2}\right)\left(1+\frac{2\Phi}{c^2}\right)\left(\delta^{ij}-h^{ij}\right),\nonumber
\ea
which simplifies greatly to
\begin{equation}
  \gamma_{ij}=h_{ij}.
  \label{eq:heqgamma}
\end{equation}
%
The exactly same result is seen in the
case of gravitational waves in a Friedmann-Robertson-Walker (FRW) background~\cite{escamilla,jimenez2017oct}.

Using Eq.~(\ref{eq:Gammadefn}), we compute the perturbed $R_{\mu\nu}(q)$:
\begin{eqnarray*}
  R_{00}(q)&=& \nabla^2\Phi_q -\Phi_q{}_{,i,j}\gamma^{ij},\\
  R_{ij}(q)&=& \frac{1}{2c^2}\left(1-\frac{4\Phi_q}{c^2}\right)\ddot{\gamma}_{ij}+\frac{1}{c^2}\nabla^2\Phi_q\left(\delta_{ij}+\gamma_{ij}\right)\nonumber\\
  &&-\frac{1}{2}\nabla^2\gamma_{ij}-\frac{1}{c^2}\Phi_q{}_{,k,l}\gamma^{kl}\delta_{ij}.
\end{eqnarray*}
From the $\mu=0$, $\nu=0$ perturbed field equation [Eq.~(\ref{eq:gammavarn})], we get $\Phi_q{}_{,i,j}\gamma^{ij} \ll \nabla^2\Phi_q$. 
Using this relation in the $\mu=i$, $\nu=j$ equation, we get
\begin{eqnarray*}
  \left(1-\frac{2\Phi_q}{c^2}\right)\left(\delta_{ij}+\gamma_{ij}\right)&=& \left(1-\frac{2\Phi}{c^2}\right)\left(\delta_{ij}+h_{ij}\right)\nonumber\\
  && \hspace{-4.0cm}+\frac{\kappa}{2c^2}\left[\left(1-\frac{4\Phi_q}{c^2}\right)\ddot{\gamma}_{ij}-c^2\nabla^2\gamma_{ij}\right]+
  \frac{\kappa}{c^2}\nabla^2\Phi_q \left(\delta_{ij}+\gamma_{ij}\right),
\end{eqnarray*}
which [after using Eqs.~(\ref{eq:linearized_eibif2}) and (\ref{eq:heqgamma})] simplifies to 
\begin{equation}
  \ddot{h}_{ij}-c^2\left(1+\frac{4\Phi_q}{c^2}\right)\nabla^2 h_{ij}=0.
  \label{eq:wave}
\end{equation}
The two polarization modes $\times$ and $+$ of the radial component of the gravitational-wave amplitude $A_{\times,+}= r h_{ij}$ satisfy the following one-dimensional wave equation:
\be
{\ddot A} + c^2 \left(1+\frac{4\Phi_q}{c^2}\right) k^2 A = 0.\label{ampEarth}
\ee
We have used $\nabla^{2}=-k^{2}$, where $k$ is the wave number. Let us take a plan wave solution of (\ref{ampEarth})
\be
A(t,\omega) = \tilde A(\omega)\,e^{i(k r - \omega t)}.
\ee
Inserting this into the wave equation (\ref{ampEarth}) gives the dispersion relation as
 \be
 - \omega^2 + c^2 \left(1+\frac{4\Phi_q}{c^2}\right) k^2 = 0.
 \ee
Thus the speed of gravitational waves in the background of Earth's gravitational field becomes
\ba
  v_{gw} \equiv \frac{d\omega}{dk} &=& c \left(1+\frac{4\Phi_q}{c^2}\right)^{1/2}\nonumber\\
  &\approx& c\left(1+\frac{2\Phi}{c^2}-\frac{4\pi G \,\kappa\, \rho}{c^2}\right),
  \label{eq:speed_gw_e}
\ea
where we used Eq.~(\ref{eq:linearized_eibif1}). Note that in GR the gravitational waves propagate with the speed of light, which in this
case is $v_{EM}\approx c\left(1+\frac{2\Phi}{c^2}\right)$. At the surface of Earth, $\Phi/c^2\approx -10^{-9}$.

{\it Bound on $\kappa$ from the speed of gravity in the Earth: } 
Cornish {\em et al.} \cite{cornish2017} obtained upper and lower bounds on the speed of gravitational wave propagation from the time delay
between gravitational-wave signals (reported by the LIGO Scientific and Virgo Collaborations) arriving at widely
separated Earth-based detectors. Their bounds are given by $0.55c<v_{gw}<1.42c$. Although this bound is very crude and will improve
with more detections and more detectors joining the worldwide network, there may be a signature of
modified gravity where the speed of gravitational waves is different from the speed of light. This feature indeedp resent in EiBI theory, and
we can use the bounds on $v_{gw}$ to put a bound on $\kappa$.
Assuming $\vert\frac{v_{gw}-c}{c} \vert <0.1$, we get $\vert\kappa\vert \lesssim 10^{21}\, m^2$, where we have used the matter density of Earth
$\rho_{Earth}=5.5\times 10^3\, kg/m^3$ in Eq.~(\ref{eq:speed_gw_e}).

\section{Speed of gravity in the background of FRW Universe}
The time delay between the recently detected gravitational-wave signal GW170817 and the associated $\gamma$-ray burst
GRB 170817A \cite{gw170817} also gives bounds on $v_{gw}$ \cite{abbot_2017oct}. The source was localized at a luminosity distance of
40 Mpc. Since the signals were propagating over an intergalactic distance, we consider the gravitational wave propagating in a background
FRW spacetime in cosmology. Thus, we solve the EiBI field equations in the FRW background for the tensor perturbations $h_{ij}$ and obtain 
the gravitational-wave propagation equation (in cosmic time) as  \cite{escamilla,jimenez2017oct}
\be
\ddot h_{ij} + \left(3H + \frac{\dot\alpha}{\alpha}\right) \dot h_{ij} - c^2 \beta^2 \frac{\nabla^{2}}{a^{2}} h_{ij} = 0\,,\label{hij_eq}
\ee
where 
\ba
\alpha = \lambda + \frac{8\pi G \,\kappa \,\rho}{c^2}\,,~~~~~~~~\beta^{2} &=& \left|\frac{ \lambda - \frac{8\pi G\, \kappa \,p}{c^4}}{\lambda + \frac{8\pi G \,\kappa \,\rho}{c^2}}\right|\,,\nonumber
\ea
and $\rho$ and $p$ are the energy density and pressure of the matter in the Universe, $H=\dot{a}/a$ is the Hubble parameter, and $a$ is the scale factor in the FRW spacetime. From the coefficient of $\dot h_{ij}$ in Eq.~(\ref{hij_eq}), we see that in addition to the cosmological damping proportional to $3H$, there is also an extra damping factor due to EiBI gravity proportional to $\frac{\dot\alpha}{\alpha}$. On subhorizon propagation distance scales this term is small.

We go to Fourier space and redefine the perturbations $h_{ij}$ as $h_{ij} = \frac{\mu_{ij}}{a\sqrt{\alpha}}$. 
The wave equation (\ref{hij_eq}) in conformal time $\tau$, defined via $d\tau = dt/a$, can be written as
\be
\mu_{ij}^{\p\p} + \left[-\frac{a^{\p\p}}{a} - \frac{\alpha^{\p\p}}{2\alpha} - \frac{a^{\p}\alpha^{\p}}{a\,\alpha}
- \frac{{\alpha^{\p}}^{2}}{4\alpha^{2}} + c^2 \beta^2 k^2 \right] \mu_{ij} = 0\,,
\ee
where prime denotes a derivative with respect to the conformal time $\tau$\,. 
On subhorizon scales we can safely ignore all of the terms inside the square brackets compared to the mode scale $k^2$, and the above equation
reduces to
\be
\mu_{ij}^{\p\p} + c^2 \beta^2 k^2 \mu_{ij} = 0.
\ee
The two polarization modes $\times$ and $+$ of the radial component of the gravitational 
wave amplitude $A_{\times,+}= r \mu_{ij}$ satisfy the following one dimensional wave equation
at large distances from the source
\be
{A^{\p\p}} + c^2 \beta^2 k^2 A = 0.\label{amp}
\ee
Let the solution of Eq. (\ref{amp}) be
\be
A(\tau,\omega) = \tilde A(\omega)\,e^{i k r - \int i \omega d\tau}.
\ee
Inserting this into the wave equation (\ref{amp}) gives the dispersion relation as
 \be
 - \omega^2 + c^2 \beta^2 k^2 = 0,
 \ee
and therefore the speed of gravitational waves can be given as
\be
v_{gw} \equiv \frac{d\omega}{dk} = c\beta = c \left|\frac{ \lambda - \frac{8\pi G \,\kappa\, p}{c^4}}
{ \lambda + \frac{8\pi G\, \kappa\, \rho}{c^2}}\right|^{1/2}\,.\label{v_GW}
\ee
The difference between the speed of gravity $v_{gw}$ and the speed of light in vacuum $c$ becomes
\be
\frac{v_{gw}-c}{c} = \left|\frac{ \lambda - \frac{8\pi G \,\kappa \,p}{c^4}}
{ \lambda + \frac{8\pi G\, \kappa\, \rho}{c^2}}\right|^{1/2}-1\,.\label{speed_diff}
\ee
 
{\it Bound on $\kappa$ from the speed of gravity in a FRW Universe: }
The observed time delay of $1.74\pm 0.05$ s between the gravity wave from the neutron star merger event and 
the light from the subsequent $\gamma$-ray burst constrain the difference between the speed of gravity and the speed of 
light to be between $-3\times 10^{-15}c$ and $+7\times 10^{-16}c$. Inserting $\left|\frac{v_{gw}-c}{c}\right|\lesssim 10^{-15}$,
the present energy density of the Universe to be at the critical density $\rho=\rho_{c0}$, the cosmological constant
$\Lambda=\frac{8\pi G \,\rho_{DE}}{c^{2}},\, \rho_{DE}\approx 0.7\rho_{c0}$, and pressure $p=-\rho_{DE}c^2$ into (\ref{speed_diff}),
we obtain $\vert\kappa\vert \lesssim 10^{37}\, m^2$.

\section{Conclusions}
In this article we demonstrated how the signature of some of the modified theories of gravity may be imprinted in the speed of gravitational waves. In particular, we put constraints on EiBI gravity from the bounds on the speed of gravity from the recent direct detection of GW signals from binary black hole and neutron star mergers. 

We derived the gravitational-wave propagation equation in the background of Earth's interior and used it to put a bound on $\kappa$. From the bound on the speed of gravity derived from the time delay of GW signals at different Earth-based detectors, we obtained  $\vert \kappa \vert \lesssim 10^{21} \, m^2 $. Similarly, from the time delay in the signals from GW170817 and GRB 170817A events, in the background of a FRW Universe, we obtained the bound $\vert \kappa \vert \lesssim 10^{37}\, m^2$. These bounds are weaker than other bounds from neutron stars, stellar evolution, nucleosynthesis, etc. However, they constitute direct constraints on $\kappa$ from observations. Also, we note that there is no dispersion of GW in matter in EiBI gravity. Future observations of the speed of GWs and their dispersion will put tighter constraints on theories of gravity beyond GR like EiBI theory.


\section*{Acknowledgments}
S.J. acknowledges Sayan Kar for useful discussions.


\bibliographystyle{apsrev4-1}
\bibliography{reference}

\begin{thebibliography}{52}%
\makeatletter
\providecommand \@ifxundefined [1]{%
 \@ifx{#1\undefined}
}%
\providecommand \@ifnum [1]{%
 \ifnum #1\expandafter \@firstoftwo
 \else \expandafter \@secondoftwo
 \fi
}%
\providecommand \@ifx [1]{%
 \ifx #1\expandafter \@firstoftwo
 \else \expandafter \@secondoftwo
 \fi
}%
\providecommand \natexlab [1]{#1}%
\providecommand \enquote  [1]{``#1''}%
\providecommand \bibnamefont  [1]{#1}%
\providecommand \bibfnamefont [1]{#1}%
\providecommand \citenamefont [1]{#1}%
\providecommand \href@noop [0]{\@secondoftwo}%
\providecommand \href [0]{\begingroup \@sanitize@url \@href}%
\providecommand \@href[1]{\@@startlink{#1}\@@href}%
\providecommand \@@href[1]{\endgroup#1\@@endlink}%
\providecommand \@sanitize@url [0]{\catcode `\\12\catcode `\$12\catcode
  `\&12\catcode `\#12\catcode `\^12\catcode `\_12\catcode `\%12\relax}%
\providecommand \@@startlink[1]{}%
\providecommand \@@endlink[0]{}%
\providecommand \url  [0]{\begingroup\@sanitize@url \@url }%
\providecommand \@url [1]{\endgroup\@href {#1}{\urlprefix }}%
\providecommand \urlprefix  [0]{URL }%
\providecommand \Eprint [0]{\href }%
\providecommand \doibase [0]{http://dx.doi.org/}%
\providecommand \selectlanguage [0]{\@gobble}%
\providecommand \bibinfo  [0]{\@secondoftwo}%
\providecommand \bibfield  [0]{\@secondoftwo}%
\providecommand \translation [1]{[#1]}%
\providecommand \BibitemOpen [0]{}%
\providecommand \bibitemStop [0]{}%
\providecommand \bibitemNoStop [0]{.\EOS\space}%
\providecommand \EOS [0]{\spacefactor3000\relax}%
\providecommand \BibitemShut  [1]{\csname bibitem#1\endcsname}%
\let\auto@bib@innerbib\@empty
\bibitem [{\citenamefont {Will}(2014)}]{Will2014}%
  \BibitemOpen
  \bibfield  {author} {\bibinfo {author} {\bibfnamefont {C.~M.}\ \bibnamefont
  {Will}},\ }\href {\doibase 10.12942/lrr-2014-4} {\bibfield  {journal}
  {\bibinfo  {journal} {Living Reviews in Relativity}\ }\textbf {\bibinfo
  {volume} {17}},\ \bibinfo {pages} {4} (\bibinfo {year} {2014})}\BibitemShut
  {NoStop}%
\bibitem [{\citenamefont {Born}\ and\ \citenamefont {Infeld}(1934)}]{born}%
  \BibitemOpen
  \bibfield  {author} {\bibinfo {author} {\bibfnamefont {M.}~\bibnamefont
  {Born}}\ and\ \bibinfo {author} {\bibfnamefont {L.}~\bibnamefont {Infeld}},\
  }\href@noop {} {\bibfield  {journal} {\bibinfo  {journal} {Proc. R. Soc. A}\
  }\textbf {\bibinfo {volume} {144}},\ \bibinfo {pages} {425} (\bibinfo {year}
  {1934})}\BibitemShut {NoStop}%
\bibitem [{\citenamefont {Deser}\ and\ \citenamefont {Gibbons}(1998)}]{desgib}%
  \BibitemOpen
  \bibfield  {author} {\bibinfo {author} {\bibfnamefont {S.}~\bibnamefont
  {Deser}}\ and\ \bibinfo {author} {\bibfnamefont {G.~W.}\ \bibnamefont
  {Gibbons}},\ }\href@noop {} {\bibfield  {journal} {\bibinfo  {journal}
  {Classical and Quantum Gravity}\ }\textbf {\bibinfo {volume} {15}},\ \bibinfo
  {pages} {L35} (\bibinfo {year} {1998})}\BibitemShut {NoStop}%
\bibitem [{\citenamefont {Eddington}(1924)}]{edd}%
  \BibitemOpen
  \bibfield  {author} {\bibinfo {author} {\bibfnamefont {A.}~\bibnamefont
  {Eddington}},\ }\href@noop {} {\emph {\bibinfo {title} {{The Mathematical
  Theory of Relativity}}}}\ (\bibinfo  {publisher} {Cambridge University
  Press},\ \bibinfo {address} {Cambridge, England},\ \bibinfo {year}
  {1924})\BibitemShut {NoStop}%
\bibitem [{\citenamefont {Vollick}(2004)}]{vollick}%
  \BibitemOpen
  \bibfield  {author} {\bibinfo {author} {\bibfnamefont {D.~N.}\ \bibnamefont
  {Vollick}},\ }\href {\doibase 10.1103/PhysRevD.69.064030} {\bibfield
  {journal} {\bibinfo  {journal} {Phys. Rev. D}\ }\textbf {\bibinfo {volume}
  {69}},\ \bibinfo {pages} {064030} (\bibinfo {year} {2004})}\BibitemShut
  {NoStop}%
\bibitem [{\citenamefont {Vollick}(2005)}]{vollick2}%
  \BibitemOpen
  \bibfield  {author} {\bibinfo {author} {\bibfnamefont {D.~N.}\ \bibnamefont
  {Vollick}},\ }\href {\doibase 10.1103/PhysRevD.72.084026} {\bibfield
  {journal} {\bibinfo  {journal} {Phys. Rev. D}\ }\textbf {\bibinfo {volume}
  {72}},\ \bibinfo {pages} {084026} (\bibinfo {year} {2005})}\BibitemShut
  {NoStop}%
\bibitem [{\citenamefont {{Vollick}}(2006)}]{vollick3}%
  \BibitemOpen
  \bibfield  {author} {\bibinfo {author} {\bibfnamefont {D.~N.}\ \bibnamefont
  {{Vollick}}},\ }\href@noop {} {\bibfield  {journal} {\bibinfo  {journal}
  {ArXiv e-prints}\ } (\bibinfo {year} {2006})},\ \Eprint
  {http://arxiv.org/abs/gr-qc/0601136} {gr-qc/0601136} \BibitemShut {NoStop}%
\bibitem [{\citenamefont {Ba\~nados}\ and\ \citenamefont
  {Ferreira}(2010)}]{banados}%
  \BibitemOpen
  \bibfield  {author} {\bibinfo {author} {\bibfnamefont {M.}~\bibnamefont
  {Ba\~nados}}\ and\ \bibinfo {author} {\bibfnamefont {P.~G.}\ \bibnamefont
  {Ferreira}},\ }\href {\doibase 10.1103/PhysRevLett.105.011101} {\bibfield
  {journal} {\bibinfo  {journal} {Phys. Rev. Lett.}\ }\textbf {\bibinfo
  {volume} {105}},\ \bibinfo {pages} {011101} (\bibinfo {year}
  {2010})}\BibitemShut {NoStop}%
\bibitem [{\citenamefont {{Beltr{\'a}n Jim{\'e}nez}}\ \emph
  {et~al.}(2018)\citenamefont {{Beltr{\'a}n Jim{\'e}nez}}, \citenamefont
  {{Heisenberg}}, \citenamefont {{Olmo}},\ and\ \citenamefont
  {{Rubiera-Garcia}}}]{jimenez2017}%
  \BibitemOpen
  \bibfield  {author} {\bibinfo {author} {\bibfnamefont {J.}~\bibnamefont
  {{Beltr{\'a}n Jim{\'e}nez}}}, \bibinfo {author} {\bibfnamefont
  {L.}~\bibnamefont {{Heisenberg}}}, \bibinfo {author} {\bibfnamefont {G.~J.}\
  \bibnamefont {{Olmo}}}, \ and\ \bibinfo {author} {\bibfnamefont
  {D.}~\bibnamefont {{Rubiera-Garcia}}},\ }\href {\doibase
  10.1016/j.physrep.2017.11.001} {\bibfield  {journal} {\bibinfo  {journal}
  {Phys. Rep.}\ }\textbf {\bibinfo {volume} {727}},\ \bibinfo {pages} {1}
  (\bibinfo {year} {2018})},\ \Eprint {http://arxiv.org/abs/1704.03351}
  {arXiv:1704.03351 [gr-qc]} \BibitemShut {NoStop}%
\bibitem [{\citenamefont {Pani}\ \emph {et~al.}(2011)\citenamefont {Pani},
  \citenamefont {Cardoso},\ and\ \citenamefont {Delsate}}]{cardoso}%
  \BibitemOpen
  \bibfield  {author} {\bibinfo {author} {\bibfnamefont {P.}~\bibnamefont
  {Pani}}, \bibinfo {author} {\bibfnamefont {V.}~\bibnamefont {Cardoso}}, \
  and\ \bibinfo {author} {\bibfnamefont {T.}~\bibnamefont {Delsate}},\ }\href
  {\doibase 10.1103/PhysRevLett.107.031101} {\bibfield  {journal} {\bibinfo
  {journal} {Phys. Rev. Lett.}\ }\textbf {\bibinfo {volume} {107}},\ \bibinfo
  {pages} {031101} (\bibinfo {year} {2011})}\BibitemShut {NoStop}%
\bibitem [{\citenamefont {Delsate}\ and\ \citenamefont
  {Steinhoff}(2012)}]{delsate}%
  \BibitemOpen
  \bibfield  {author} {\bibinfo {author} {\bibfnamefont {T.}~\bibnamefont
  {Delsate}}\ and\ \bibinfo {author} {\bibfnamefont {J.}~\bibnamefont
  {Steinhoff}},\ }\href {\doibase 10.1103/PhysRevLett.109.021101} {\bibfield
  {journal} {\bibinfo  {journal} {Phys. Rev. Lett.}\ }\textbf {\bibinfo
  {volume} {109}},\ \bibinfo {pages} {021101} (\bibinfo {year}
  {2012})}\BibitemShut {NoStop}%
\bibitem [{\citenamefont {Pani}\ and\ \citenamefont {Sotiriou}(2012)}]{pani}%
  \BibitemOpen
  \bibfield  {author} {\bibinfo {author} {\bibfnamefont {P.}~\bibnamefont
  {Pani}}\ and\ \bibinfo {author} {\bibfnamefont {T.~P.}\ \bibnamefont
  {Sotiriou}},\ }\href {\doibase 10.1103/PhysRevLett.109.251102} {\bibfield
  {journal} {\bibinfo  {journal} {Phys. Rev. Lett.}\ }\textbf {\bibinfo
  {volume} {109}},\ \bibinfo {pages} {251102} (\bibinfo {year}
  {2012})}\BibitemShut {NoStop}%
\bibitem [{\citenamefont {Scargill}\ \emph {et~al.}(2012)\citenamefont
  {Scargill}, \citenamefont {Banados},\ and\ \citenamefont
  {Ferreira}}]{scargill}%
  \BibitemOpen
  \bibfield  {author} {\bibinfo {author} {\bibfnamefont {J.~H.~C.}\
  \bibnamefont {Scargill}}, \bibinfo {author} {\bibfnamefont {M.}~\bibnamefont
  {Banados}}, \ and\ \bibinfo {author} {\bibfnamefont {P.~G.}\ \bibnamefont
  {Ferreira}},\ }\href {\doibase 10.1103/PhysRevD.86.103533} {\bibfield
  {journal} {\bibinfo  {journal} {Phys. Rev. D}\ }\textbf {\bibinfo {volume}
  {86}},\ \bibinfo {pages} {103533} (\bibinfo {year} {2012})}\BibitemShut
  {NoStop}%
\bibitem [{\citenamefont {Cho}\ \emph {et~al.}(2012)\citenamefont {Cho},
  \citenamefont {Kim},\ and\ \citenamefont {Moon}}]{cho}%
  \BibitemOpen
  \bibfield  {author} {\bibinfo {author} {\bibfnamefont {I.}~\bibnamefont
  {Cho}}, \bibinfo {author} {\bibfnamefont {H.-C.}\ \bibnamefont {Kim}}, \ and\
  \bibinfo {author} {\bibfnamefont {T.}~\bibnamefont {Moon}},\ }\href {\doibase
  10.1103/PhysRevD.86.084018} {\bibfield  {journal} {\bibinfo  {journal} {Phys.
  Rev. D}\ }\textbf {\bibinfo {volume} {86}},\ \bibinfo {pages} {084018}
  (\bibinfo {year} {2012})}\BibitemShut {NoStop}%
\bibitem [{\citenamefont {Cho}\ \emph {et~al.}(2013)\citenamefont {Cho},
  \citenamefont {Kim},\ and\ \citenamefont {Moon}}]{chokim}%
  \BibitemOpen
  \bibfield  {author} {\bibinfo {author} {\bibfnamefont {I.}~\bibnamefont
  {Cho}}, \bibinfo {author} {\bibfnamefont {H.-C.}\ \bibnamefont {Kim}}, \ and\
  \bibinfo {author} {\bibfnamefont {T.}~\bibnamefont {Moon}},\ }\href {\doibase
  10.1103/PhysRevLett.111.071301} {\bibfield  {journal} {\bibinfo  {journal}
  {Phys. Rev. Lett.}\ }\textbf {\bibinfo {volume} {111}},\ \bibinfo {pages}
  {071301} (\bibinfo {year} {2013})}\BibitemShut {NoStop}%
\bibitem [{\citenamefont {Escamilla-Rivera}\ \emph {et~al.}(2012)\citenamefont
  {Escamilla-Rivera}, \citenamefont {Banados},\ and\ \citenamefont
  {Ferreira}}]{escamilla}%
  \BibitemOpen
  \bibfield  {author} {\bibinfo {author} {\bibfnamefont {C.}~\bibnamefont
  {Escamilla-Rivera}}, \bibinfo {author} {\bibfnamefont {M.}~\bibnamefont
  {Banados}}, \ and\ \bibinfo {author} {\bibfnamefont {P.~G.}\ \bibnamefont
  {Ferreira}},\ }\href {\doibase 10.1103/PhysRevD.85.087302} {\bibfield
  {journal} {\bibinfo  {journal} {Phys. Rev. D}\ }\textbf {\bibinfo {volume}
  {85}},\ \bibinfo {pages} {087302} (\bibinfo {year} {2012})}\BibitemShut
  {NoStop}%
\bibitem [{\citenamefont {Yang}\ \emph {et~al.}(2013)\citenamefont {Yang},
  \citenamefont {Du},\ and\ \citenamefont {Liu}}]{linear.perturbation}%
  \BibitemOpen
  \bibfield  {author} {\bibinfo {author} {\bibfnamefont {K.}~\bibnamefont
  {Yang}}, \bibinfo {author} {\bibfnamefont {X.-L.}\ \bibnamefont {Du}}, \ and\
  \bibinfo {author} {\bibfnamefont {Y.-X.}\ \bibnamefont {Liu}},\ }\href
  {\doibase 10.1103/PhysRevD.88.124037} {\bibfield  {journal} {\bibinfo
  {journal} {Phys. Rev. D}\ }\textbf {\bibinfo {volume} {88}},\ \bibinfo
  {pages} {124037} (\bibinfo {year} {2013})}\BibitemShut {NoStop}%
\bibitem [{\citenamefont {{Wei}}\ \emph {et~al.}(2015)\citenamefont {{Wei}},
  \citenamefont {{Yang}},\ and\ \citenamefont {{Liu}}}]{wei}%
  \BibitemOpen
  \bibfield  {author} {\bibinfo {author} {\bibfnamefont {S.-W.}\ \bibnamefont
  {{Wei}}}, \bibinfo {author} {\bibfnamefont {K.}~\bibnamefont {{Yang}}}, \
  and\ \bibinfo {author} {\bibfnamefont {Y.-X.}\ \bibnamefont {{Liu}}},\ }\href
  {\doibase 10.1140/epjc/s10052-015-3469-7} {\bibfield  {journal} {\bibinfo
  {journal} {European Physical Journal C}\ }\textbf {\bibinfo {volume} {75}},\
  \bibinfo {eid} {253} (\bibinfo {year} {2015})},\ \Eprint
  {http://arxiv.org/abs/1405.2178} {arXiv:1405.2178 [gr-qc]} \BibitemShut
  {NoStop}%
\bibitem [{\citenamefont {Jana}\ and\ \citenamefont {Kar}(2013)}]{jana}%
  \BibitemOpen
  \bibfield  {author} {\bibinfo {author} {\bibfnamefont {S.}~\bibnamefont
  {Jana}}\ and\ \bibinfo {author} {\bibfnamefont {S.}~\bibnamefont {Kar}},\
  }\href {\doibase 10.1103/PhysRevD.88.024013} {\bibfield  {journal} {\bibinfo
  {journal} {Phys. Rev. D}\ }\textbf {\bibinfo {volume} {88}},\ \bibinfo
  {pages} {024013} (\bibinfo {year} {2013})}\BibitemShut {NoStop}%
\bibitem [{\citenamefont {Jana}\ and\ \citenamefont {Kar}(2015)}]{jana2}%
  \BibitemOpen
  \bibfield  {author} {\bibinfo {author} {\bibfnamefont {S.}~\bibnamefont
  {Jana}}\ and\ \bibinfo {author} {\bibfnamefont {S.}~\bibnamefont {Kar}},\
  }\href {\doibase 10.1103/PhysRevD.92.084004} {\bibfield  {journal} {\bibinfo
  {journal} {Phys. Rev. D}\ }\textbf {\bibinfo {volume} {92}},\ \bibinfo
  {pages} {084004} (\bibinfo {year} {2015})}\BibitemShut {NoStop}%
\bibitem [{\citenamefont {Jana}\ and\ \citenamefont {Kar}(2016)}]{jana4}%
  \BibitemOpen
  \bibfield  {author} {\bibinfo {author} {\bibfnamefont {S.}~\bibnamefont
  {Jana}}\ and\ \bibinfo {author} {\bibfnamefont {S.}~\bibnamefont {Kar}},\
  }\href {\doibase 10.1103/PhysRevD.94.064016} {\bibfield  {journal} {\bibinfo
  {journal} {Phys. Rev. D}\ }\textbf {\bibinfo {volume} {94}},\ \bibinfo
  {pages} {064016} (\bibinfo {year} {2016})}\BibitemShut {NoStop}%
\bibitem [{\citenamefont {{Jana}}\ and\ \citenamefont
  {{Kar}}(2017)}]{jana2017}%
  \BibitemOpen
  \bibfield  {author} {\bibinfo {author} {\bibfnamefont {S.}~\bibnamefont
  {{Jana}}}\ and\ \bibinfo {author} {\bibfnamefont {S.}~\bibnamefont {{Kar}}},\
  }\href {\doibase 10.1103/PhysRevD.96.024050} {\bibfield  {journal} {\bibinfo
  {journal} {Phys. Rev. D}\ }\textbf {\bibinfo {volume} {96}},\ \bibinfo {eid}
  {024050} (\bibinfo {year} {2017})},\ \Eprint
  {http://arxiv.org/abs/1706.03209} {arXiv:1706.03209 [gr-qc]} \BibitemShut
  {NoStop}%
\bibitem [{\citenamefont {Shaikh}(2015)}]{rajibul}%
  \BibitemOpen
  \bibfield  {author} {\bibinfo {author} {\bibfnamefont {R.}~\bibnamefont
  {Shaikh}},\ }\href {\doibase 10.1103/PhysRevD.92.024015} {\bibfield
  {journal} {\bibinfo  {journal} {Phys. Rev. D}\ }\textbf {\bibinfo {volume}
  {92}},\ \bibinfo {pages} {024015} (\bibinfo {year} {2015})}\BibitemShut
  {NoStop}%
\bibitem [{\citenamefont {Sotani}\ and\ \citenamefont
  {Miyamoto}(2014)}]{sotani}%
  \BibitemOpen
  \bibfield  {author} {\bibinfo {author} {\bibfnamefont {H.}~\bibnamefont
  {Sotani}}\ and\ \bibinfo {author} {\bibfnamefont {U.}~\bibnamefont
  {Miyamoto}},\ }\href {\doibase 10.1103/PhysRevD.90.124087} {\bibfield
  {journal} {\bibinfo  {journal} {Phys. Rev. D}\ }\textbf {\bibinfo {volume}
  {90}},\ \bibinfo {pages} {124087} (\bibinfo {year} {2014})}\BibitemShut
  {NoStop}%
\bibitem [{\citenamefont {Olmo}\ \emph {et~al.}(2014)\citenamefont {Olmo},
  \citenamefont {Rubiera-Garcia},\ and\ \citenamefont
  {Sanchis-Alepuz}}]{eibiwormhole}%
  \BibitemOpen
  \bibfield  {author} {\bibinfo {author} {\bibfnamefont {G.~J.}\ \bibnamefont
  {Olmo}}, \bibinfo {author} {\bibfnamefont {D.}~\bibnamefont
  {Rubiera-Garcia}}, \ and\ \bibinfo {author} {\bibfnamefont {H.}~\bibnamefont
  {Sanchis-Alepuz}},\ }\href {\doibase 10.1140/epjc/s10052-014-2804-8}
  {\bibfield  {journal} {\bibinfo  {journal} {The European Physical Journal C}\
  }\textbf {\bibinfo {volume} {74}},\ \bibinfo {pages} {2804} (\bibinfo {year}
  {2014})}\BibitemShut {NoStop}%
\bibitem [{\citenamefont {Bazeia}\ \emph {et~al.}(2017)\citenamefont {Bazeia},
  \citenamefont {Losano}, \citenamefont {Olmo},\ and\ \citenamefont
  {Rubiera-Garcia}}]{BTZ_typesoln}%
  \BibitemOpen
  \bibfield  {author} {\bibinfo {author} {\bibfnamefont {D.}~\bibnamefont
  {Bazeia}}, \bibinfo {author} {\bibfnamefont {L.}~\bibnamefont {Losano}},
  \bibinfo {author} {\bibfnamefont {G.~J.}\ \bibnamefont {Olmo}}, \ and\
  \bibinfo {author} {\bibfnamefont {D.}~\bibnamefont {Rubiera-Garcia}},\
  }\href@noop {} {\bibfield  {journal} {\bibinfo  {journal} {Classical and
  Quantum Gravity}\ }\textbf {\bibinfo {volume} {34}},\ \bibinfo {pages}
  {045006} (\bibinfo {year} {2017})}\BibitemShut {NoStop}%
\bibitem [{\citenamefont {Casanellas}\ \emph {et~al.}(2012)\citenamefont
  {Casanellas}, \citenamefont {Pani}, \citenamefont {Lopes},\ and\
  \citenamefont {Cardoso}}]{casanellas}%
  \BibitemOpen
  \bibfield  {author} {\bibinfo {author} {\bibfnamefont {J.}~\bibnamefont
  {Casanellas}}, \bibinfo {author} {\bibfnamefont {P.}~\bibnamefont {Pani}},
  \bibinfo {author} {\bibfnamefont {I.}~\bibnamefont {Lopes}}, \ and\ \bibinfo
  {author} {\bibfnamefont {V.}~\bibnamefont {Cardoso}},\ }\href@noop {}
  {\bibfield  {journal} {\bibinfo  {journal} {The Astrophysical Journal}\
  }\textbf {\bibinfo {volume} {745}},\ \bibinfo {pages} {15} (\bibinfo {year}
  {2012})}\BibitemShut {NoStop}%
\bibitem [{\citenamefont {Avelino}(2012{\natexlab{a}})}]{avelino}%
  \BibitemOpen
  \bibfield  {author} {\bibinfo {author} {\bibfnamefont {P.~P.}\ \bibnamefont
  {Avelino}},\ }\href {\doibase 10.1103/PhysRevD.85.104053} {\bibfield
  {journal} {\bibinfo  {journal} {Phys. Rev. D}\ }\textbf {\bibinfo {volume}
  {85}},\ \bibinfo {pages} {104053} (\bibinfo {year}
  {2012}{\natexlab{a}})}\BibitemShut {NoStop}%
\bibitem [{\citenamefont {Sham}\ \emph {et~al.}(2012)\citenamefont {Sham},
  \citenamefont {Lin},\ and\ \citenamefont {Leung}}]{sham}%
  \BibitemOpen
  \bibfield  {author} {\bibinfo {author} {\bibfnamefont {Y.-H.}\ \bibnamefont
  {Sham}}, \bibinfo {author} {\bibfnamefont {L.-M.}\ \bibnamefont {Lin}}, \
  and\ \bibinfo {author} {\bibfnamefont {P.~T.}\ \bibnamefont {Leung}},\ }\href
  {\doibase 10.1103/PhysRevD.86.064015} {\bibfield  {journal} {\bibinfo
  {journal} {Phys. Rev. D}\ }\textbf {\bibinfo {volume} {86}},\ \bibinfo
  {pages} {064015} (\bibinfo {year} {2012})}\BibitemShut {NoStop}%
\bibitem [{\citenamefont {Sham}\ \emph {et~al.}(2013)\citenamefont {Sham},
  \citenamefont {Leung},\ and\ \citenamefont {Lin}}]{sham2}%
  \BibitemOpen
  \bibfield  {author} {\bibinfo {author} {\bibfnamefont {Y.-H.}\ \bibnamefont
  {Sham}}, \bibinfo {author} {\bibfnamefont {P.~T.}\ \bibnamefont {Leung}}, \
  and\ \bibinfo {author} {\bibfnamefont {L.-M.}\ \bibnamefont {Lin}},\ }\href
  {\doibase 10.1103/PhysRevD.87.061503} {\bibfield  {journal} {\bibinfo
  {journal} {Phys. Rev. D}\ }\textbf {\bibinfo {volume} {87}},\ \bibinfo
  {pages} {061503} (\bibinfo {year} {2013})}\BibitemShut {NoStop}%
\bibitem [{\citenamefont {Harko}\ \emph {et~al.}(2013)\citenamefont {Harko},
  \citenamefont {Lobo}, \citenamefont {Mak},\ and\ \citenamefont
  {Sushkov}}]{structure.exotic.star}%
  \BibitemOpen
  \bibfield  {author} {\bibinfo {author} {\bibfnamefont {T.}~\bibnamefont
  {Harko}}, \bibinfo {author} {\bibfnamefont {F.~S.~N.}\ \bibnamefont {Lobo}},
  \bibinfo {author} {\bibfnamefont {M.~K.}\ \bibnamefont {Mak}}, \ and\
  \bibinfo {author} {\bibfnamefont {S.~V.}\ \bibnamefont {Sushkov}},\ }\href
  {\doibase 10.1103/PhysRevD.88.044032} {\bibfield  {journal} {\bibinfo
  {journal} {Phys. Rev. D}\ }\textbf {\bibinfo {volume} {88}},\ \bibinfo
  {pages} {044032} (\bibinfo {year} {2013})}\BibitemShut {NoStop}%
\bibitem [{\citenamefont {Sotani}(2014{\natexlab{a}})}]{sotani.neutron.star}%
  \BibitemOpen
  \bibfield  {author} {\bibinfo {author} {\bibfnamefont {H.}~\bibnamefont
  {Sotani}},\ }\href {\doibase 10.1103/PhysRevD.89.104005} {\bibfield
  {journal} {\bibinfo  {journal} {Phys. Rev. D}\ }\textbf {\bibinfo {volume}
  {89}},\ \bibinfo {pages} {104005} (\bibinfo {year}
  {2014}{\natexlab{a}})}\BibitemShut {NoStop}%
\bibitem [{\citenamefont
  {Sotani}(2014{\natexlab{b}})}]{sotani.stellar.oscillations}%
  \BibitemOpen
  \bibfield  {author} {\bibinfo {author} {\bibfnamefont {H.}~\bibnamefont
  {Sotani}},\ }\href {\doibase 10.1103/PhysRevD.89.124037} {\bibfield
  {journal} {\bibinfo  {journal} {Phys. Rev. D}\ }\textbf {\bibinfo {volume}
  {89}},\ \bibinfo {pages} {124037} (\bibinfo {year}
  {2014}{\natexlab{b}})}\BibitemShut {NoStop}%
\bibitem [{\citenamefont {{Sotani}}(2015)}]{sotani.magnetic.star}%
  \BibitemOpen
  \bibfield  {author} {\bibinfo {author} {\bibfnamefont {H.}~\bibnamefont
  {{Sotani}}},\ }\href {\doibase 10.1103/PhysRevD.91.084020} {\bibfield
  {journal} {\bibinfo  {journal} {Phys. Rev. D}\ }\textbf {\bibinfo {volume}
  {91}},\ \bibinfo {eid} {084020} (\bibinfo {year} {2015})},\ \Eprint
  {http://arxiv.org/abs/1503.07942} {arXiv:1503.07942 [astro-ph.HE]}
  \BibitemShut {NoStop}%
\bibitem [{\citenamefont {Odintsov}\ \emph {et~al.}(2014)\citenamefont
  {Odintsov}, \citenamefont {Olmo},\ and\ \citenamefont
  {Rubiera-Garcia}}]{odintsov}%
  \BibitemOpen
  \bibfield  {author} {\bibinfo {author} {\bibfnamefont {S.~D.}\ \bibnamefont
  {Odintsov}}, \bibinfo {author} {\bibfnamefont {G.~J.}\ \bibnamefont {Olmo}},
  \ and\ \bibinfo {author} {\bibfnamefont {D.}~\bibnamefont {Rubiera-Garcia}},\
  }\href {\doibase 10.1103/PhysRevD.90.044003} {\bibfield  {journal} {\bibinfo
  {journal} {Phys. Rev. D}\ }\textbf {\bibinfo {volume} {90}},\ \bibinfo
  {pages} {044003} (\bibinfo {year} {2014})}\BibitemShut {NoStop}%
\bibitem [{\citenamefont {Fernandes}\ and\ \citenamefont
  {Lahiri}(2015)}]{fernandes}%
  \BibitemOpen
  \bibfield  {author} {\bibinfo {author} {\bibfnamefont {K.}~\bibnamefont
  {Fernandes}}\ and\ \bibinfo {author} {\bibfnamefont {A.}~\bibnamefont
  {Lahiri}},\ }\href {\doibase 10.1103/PhysRevD.91.044014} {\bibfield
  {journal} {\bibinfo  {journal} {Phys. Rev. D}\ }\textbf {\bibinfo {volume}
  {91}},\ \bibinfo {pages} {044014} (\bibinfo {year} {2015})}\BibitemShut
  {NoStop}%
\bibitem [{\citenamefont {Delhom-Latorre}\ \emph {et~al.}(2018)\citenamefont
  {Delhom-Latorre}, \citenamefont {Olmo},\ and\ \citenamefont
  {Ronco}}]{latorre2017}%
  \BibitemOpen
  \bibfield  {author} {\bibinfo {author} {\bibfnamefont {A.}~\bibnamefont
  {Delhom-Latorre}}, \bibinfo {author} {\bibfnamefont {G.~J.}\ \bibnamefont
  {Olmo}}, \ and\ \bibinfo {author} {\bibfnamefont {M.}~\bibnamefont {Ronco}},\
  }\href {\doibase https://doi.org/10.1016/j.physletb.2018.03.002} {\bibfield
  {journal} {\bibinfo  {journal} {Physics Letters B}\ }\textbf {\bibinfo
  {volume} {780}},\ \bibinfo {pages} {294 } (\bibinfo {year}
  {2018})}\BibitemShut {NoStop}%
\bibitem [{\citenamefont {{Abbott}}\ \emph {et~al.}(2016)\citenamefont
  {{Abbott}}, \citenamefont {{Abbott}}, \citenamefont {{Abbott}}, \citenamefont
  {{Abernathy}}, \citenamefont {{Acernese}}, \citenamefont {{Ackley}},
  \citenamefont {{Adams}}, \citenamefont {{Adams}}, \citenamefont {{Addesso}},
  \citenamefont {{Adhikari}},\ and\ \citenamefont {et~al.}}]{abbott}%
  \BibitemOpen
  \bibfield  {author} {\bibinfo {author} {\bibfnamefont {B.~P.}\ \bibnamefont
  {{Abbott}}}, \bibinfo {author} {\bibfnamefont {R.}~\bibnamefont {{Abbott}}},
  \bibinfo {author} {\bibfnamefont {T.~D.}\ \bibnamefont {{Abbott}}}, \bibinfo
  {author} {\bibfnamefont {M.~R.}\ \bibnamefont {{Abernathy}}}, \bibinfo
  {author} {\bibfnamefont {F.}~\bibnamefont {{Acernese}}}, \bibinfo {author}
  {\bibfnamefont {K.}~\bibnamefont {{Ackley}}}, \bibinfo {author}
  {\bibfnamefont {C.}~\bibnamefont {{Adams}}}, \bibinfo {author} {\bibfnamefont
  {T.}~\bibnamefont {{Adams}}}, \bibinfo {author} {\bibfnamefont
  {P.}~\bibnamefont {{Addesso}}}, \bibinfo {author} {\bibfnamefont {R.~X.}\
  \bibnamefont {{Adhikari}}}, \ and\ \bibinfo {author} {\bibnamefont
  {et~al.}},\ }\href {\doibase 10.1103/PhysRevLett.116.221101} {\bibfield
  {journal} {\bibinfo  {journal} {Physical Review Letters}\ }\textbf {\bibinfo
  {volume} {116}},\ \bibinfo {eid} {221101} (\bibinfo {year} {2016})},\ \Eprint
  {http://arxiv.org/abs/1602.03841} {arXiv:1602.03841 [gr-qc]} \BibitemShut
  {NoStop}%
\bibitem [{\citenamefont {{Abbott}}\ \emph
  {et~al.}(2017{\natexlab{a}})\citenamefont {{Abbott}}, \citenamefont
  {{Abbott}}, \citenamefont {{Abbott}}, \citenamefont {{Acernese}},
  \citenamefont {{Ackley}}, \citenamefont {{Adams}}, \citenamefont {{Adams}},
  \citenamefont {{Addesso}}, \citenamefont {{Adhikari}}, \citenamefont
  {{Adya}},\ and\ \citenamefont {et~al.}}]{gw170817}%
  \BibitemOpen
  \bibfield  {author} {\bibinfo {author} {\bibfnamefont {B.~P.}\ \bibnamefont
  {{Abbott}}}, \bibinfo {author} {\bibfnamefont {R.}~\bibnamefont {{Abbott}}},
  \bibinfo {author} {\bibfnamefont {T.~D.}\ \bibnamefont {{Abbott}}}, \bibinfo
  {author} {\bibfnamefont {F.}~\bibnamefont {{Acernese}}}, \bibinfo {author}
  {\bibfnamefont {K.}~\bibnamefont {{Ackley}}}, \bibinfo {author}
  {\bibfnamefont {C.}~\bibnamefont {{Adams}}}, \bibinfo {author} {\bibfnamefont
  {T.}~\bibnamefont {{Adams}}}, \bibinfo {author} {\bibfnamefont
  {P.}~\bibnamefont {{Addesso}}}, \bibinfo {author} {\bibfnamefont {R.~X.}\
  \bibnamefont {{Adhikari}}}, \bibinfo {author} {\bibfnamefont {V.~B.}\
  \bibnamefont {{Adya}}}, \ and\ \bibinfo {author} {\bibnamefont {et~al.}},\
  }\href {\doibase 10.1103/PhysRevLett.119.161101} {\bibfield  {journal}
  {\bibinfo  {journal} {Physical Review Letters}\ }\textbf {\bibinfo {volume}
  {119}},\ \bibinfo {eid} {161101} (\bibinfo {year} {2017}{\natexlab{a}})},\
  \Eprint {http://arxiv.org/abs/1710.05832} {arXiv:1710.05832 [gr-qc]}
  \BibitemShut {NoStop}%
\bibitem [{\citenamefont {{Abbott}}\ \emph
  {et~al.}(2017{\natexlab{b}})\citenamefont {{Abbott}}, \citenamefont
  {{Abbott}}, \citenamefont {{Abbott}}, \citenamefont {{Acernese}},
  \citenamefont {{Ackley}}, \citenamefont {{Adams}}, \citenamefont {{Adams}},
  \citenamefont {{Addesso}}, \citenamefont {{Adhikari}}, \citenamefont
  {{Adya}},\ and\ \citenamefont {et~al.}}]{abbot_2017oct}%
  \BibitemOpen
  \bibfield  {author} {\bibinfo {author} {\bibfnamefont {B.~P.}\ \bibnamefont
  {{Abbott}}}, \bibinfo {author} {\bibfnamefont {R.}~\bibnamefont {{Abbott}}},
  \bibinfo {author} {\bibfnamefont {T.~D.}\ \bibnamefont {{Abbott}}}, \bibinfo
  {author} {\bibfnamefont {F.}~\bibnamefont {{Acernese}}}, \bibinfo {author}
  {\bibfnamefont {K.}~\bibnamefont {{Ackley}}}, \bibinfo {author}
  {\bibfnamefont {C.}~\bibnamefont {{Adams}}}, \bibinfo {author} {\bibfnamefont
  {T.}~\bibnamefont {{Adams}}}, \bibinfo {author} {\bibfnamefont
  {P.}~\bibnamefont {{Addesso}}}, \bibinfo {author} {\bibfnamefont {R.~X.}\
  \bibnamefont {{Adhikari}}}, \bibinfo {author} {\bibfnamefont {V.~B.}\
  \bibnamefont {{Adya}}}, \ and\ \bibinfo {author} {\bibnamefont {et~al.}},\
  }\href {\doibase 10.3847/2041-8213/aa920c} {\bibfield  {journal} {\bibinfo
  {journal} {The Astrophysical Journal Letters}\ }\textbf {\bibinfo {volume}
  {848}},\ \bibinfo {eid} {L13} (\bibinfo {year} {2017}{\natexlab{b}})},\
  \Eprint {http://arxiv.org/abs/1710.05834} {arXiv:1710.05834 [astro-ph.HE]}
  \BibitemShut {NoStop}%
\bibitem [{\citenamefont {Cornish}\ \emph {et~al.}(2017)\citenamefont
  {Cornish}, \citenamefont {Blas},\ and\ \citenamefont
  {Nardini}}]{cornish2017}%
  \BibitemOpen
  \bibfield  {author} {\bibinfo {author} {\bibfnamefont {N.}~\bibnamefont
  {Cornish}}, \bibinfo {author} {\bibfnamefont {D.}~\bibnamefont {Blas}}, \
  and\ \bibinfo {author} {\bibfnamefont {G.}~\bibnamefont {Nardini}},\ }\href
  {\doibase 10.1103/PhysRevLett.119.161102} {\bibfield  {journal} {\bibinfo
  {journal} {Phys. Rev. Lett.}\ }\textbf {\bibinfo {volume} {119}},\ \bibinfo
  {pages} {161102} (\bibinfo {year} {2017})}\BibitemShut {NoStop}%
\bibitem [{\citenamefont {{Lombriser}}\ and\ \citenamefont
  {{Taylor}}(2016)}]{lombriser2016}%
  \BibitemOpen
  \bibfield  {author} {\bibinfo {author} {\bibfnamefont {L.}~\bibnamefont
  {{Lombriser}}}\ and\ \bibinfo {author} {\bibfnamefont {A.}~\bibnamefont
  {{Taylor}}},\ }\href {\doibase 10.1088/1475-7516/2016/03/031} {\bibfield
  {journal} {\bibinfo  {journal} {JCAP}\ }\textbf {\bibinfo {volume} {3}},\
  \bibinfo {eid} {031} (\bibinfo {year} {2016})},\ \Eprint
  {http://arxiv.org/abs/1509.08458} {arXiv:1509.08458} \BibitemShut {NoStop}%
\bibitem [{\citenamefont {{Lombriser}}\ and\ \citenamefont
  {{Lima}}(2017)}]{lombriser2017}%
  \BibitemOpen
  \bibfield  {author} {\bibinfo {author} {\bibfnamefont {L.}~\bibnamefont
  {{Lombriser}}}\ and\ \bibinfo {author} {\bibfnamefont {N.~A.}\ \bibnamefont
  {{Lima}}},\ }\href {\doibase 10.1016/j.physletb.2016.12.048} {\bibfield
  {journal} {\bibinfo  {journal} {Physics Letters B}\ }\textbf {\bibinfo
  {volume} {765}},\ \bibinfo {pages} {382} (\bibinfo {year} {2017})},\ \Eprint
  {http://arxiv.org/abs/1602.07670} {arXiv:1602.07670} \BibitemShut {NoStop}%
\bibitem [{\citenamefont {{Bettoni}}\ \emph {et~al.}(2017)\citenamefont
  {{Bettoni}}, \citenamefont {{Ezquiaga}}, \citenamefont {{Hinterbichler}},\
  and\ \citenamefont {{Zumalac{\'a}rregui}}}]{Zumalacarregui2016}%
  \BibitemOpen
  \bibfield  {author} {\bibinfo {author} {\bibfnamefont {D.}~\bibnamefont
  {{Bettoni}}}, \bibinfo {author} {\bibfnamefont {J.~M.}\ \bibnamefont
  {{Ezquiaga}}}, \bibinfo {author} {\bibfnamefont {K.}~\bibnamefont
  {{Hinterbichler}}}, \ and\ \bibinfo {author} {\bibfnamefont {M.}~\bibnamefont
  {{Zumalac{\'a}rregui}}},\ }\href {\doibase 10.1103/PhysRevD.95.084029}
  {\bibfield  {journal} {\bibinfo  {journal} {Phys. Rev. D}\ }\textbf {\bibinfo
  {volume} {95}},\ \bibinfo {eid} {084029} (\bibinfo {year} {2017})},\ \Eprint
  {http://arxiv.org/abs/1608.01982} {arXiv:1608.01982 [gr-qc]} \BibitemShut
  {NoStop}%
\bibitem [{\citenamefont {{Shoemaker}}\ and\ \citenamefont
  {{Murase}}(2017)}]{shoemaker2017}%
  \BibitemOpen
  \bibfield  {author} {\bibinfo {author} {\bibfnamefont {I.~M.}\ \bibnamefont
  {{Shoemaker}}}\ and\ \bibinfo {author} {\bibfnamefont {K.}~\bibnamefont
  {{Murase}}},\ }\href@noop {} {\bibfield  {journal} {\bibinfo  {journal}
  {ArXiv e-prints}\ } (\bibinfo {year} {2017})},\ \Eprint
  {http://arxiv.org/abs/1710.06427} {arXiv:1710.06427 [astro-ph.HE]}
  \BibitemShut {NoStop}%
\bibitem [{\citenamefont {{Baker}}\ \emph {et~al.}(2017)\citenamefont
  {{Baker}}, \citenamefont {{Bellini}}, \citenamefont {{Ferreira}},
  \citenamefont {{Lagos}}, \citenamefont {{Noller}},\ and\ \citenamefont
  {{Sawicki}}}]{baker1710}%
  \BibitemOpen
  \bibfield  {author} {\bibinfo {author} {\bibfnamefont {T.}~\bibnamefont
  {{Baker}}}, \bibinfo {author} {\bibfnamefont {E.}~\bibnamefont {{Bellini}}},
  \bibinfo {author} {\bibfnamefont {P.~G.}\ \bibnamefont {{Ferreira}}},
  \bibinfo {author} {\bibfnamefont {M.}~\bibnamefont {{Lagos}}}, \bibinfo
  {author} {\bibfnamefont {J.}~\bibnamefont {{Noller}}}, \ and\ \bibinfo
  {author} {\bibfnamefont {I.}~\bibnamefont {{Sawicki}}},\ }\href {\doibase
  10.1103/PhysRevLett.119.251301} {\bibfield  {journal} {\bibinfo  {journal}
  {Physical Review Letters}\ }\textbf {\bibinfo {volume} {119}},\ \bibinfo
  {eid} {251301} (\bibinfo {year} {2017})},\ \Eprint
  {http://arxiv.org/abs/1710.06394} {arXiv:1710.06394} \BibitemShut {NoStop}%
\bibitem [{\citenamefont {{Sakstein}}\ and\ \citenamefont
  {{Jain}}(2017)}]{Sakstein:2017xjx}%
  \BibitemOpen
  \bibfield  {author} {\bibinfo {author} {\bibfnamefont {J.}~\bibnamefont
  {{Sakstein}}}\ and\ \bibinfo {author} {\bibfnamefont {B.}~\bibnamefont
  {{Jain}}},\ }\href {\doibase 10.1103/PhysRevLett.119.251303} {\bibfield
  {journal} {\bibinfo  {journal} {Physical Review Letters}\ }\textbf {\bibinfo
  {volume} {119}},\ \bibinfo {eid} {251303} (\bibinfo {year} {2017})},\ \Eprint
  {http://arxiv.org/abs/1710.05893} {arXiv:1710.05893} \BibitemShut {NoStop}%
\bibitem [{\citenamefont {{Creminelli}}\ and\ \citenamefont
  {{Vernizzi}}(2017)}]{Creminelli:2017sry}%
  \BibitemOpen
  \bibfield  {author} {\bibinfo {author} {\bibfnamefont {P.}~\bibnamefont
  {{Creminelli}}}\ and\ \bibinfo {author} {\bibfnamefont {F.}~\bibnamefont
  {{Vernizzi}}},\ }\href {\doibase 10.1103/PhysRevLett.119.251302} {\bibfield
  {journal} {\bibinfo  {journal} {Physical Review Letters}\ }\textbf {\bibinfo
  {volume} {119}},\ \bibinfo {eid} {251302} (\bibinfo {year} {2017})},\ \Eprint
  {http://arxiv.org/abs/1710.05877} {arXiv:1710.05877} \BibitemShut {NoStop}%
\bibitem [{\citenamefont {Ezquiaga}\ and\ \citenamefont
  {Zumalac\'arregui}(2017)}]{Ezquiaga:2017ekz}%
  \BibitemOpen
  \bibfield  {author} {\bibinfo {author} {\bibfnamefont {J.~M.}\ \bibnamefont
  {Ezquiaga}}\ and\ \bibinfo {author} {\bibfnamefont {M.}~\bibnamefont
  {Zumalac\'arregui}},\ }\href {\doibase 10.1103/PhysRevLett.119.251304}
  {\bibfield  {journal} {\bibinfo  {journal} {Phys. Rev. Lett.}\ }\textbf
  {\bibinfo {volume} {119}},\ \bibinfo {pages} {251304} (\bibinfo {year}
  {2017})}\BibitemShut {NoStop}%
\bibitem [{\citenamefont {{Nojiri}}\ and\ \citenamefont
  {{Odintsov}}(2018)}]{1711.00492}%
  \BibitemOpen
  \bibfield  {author} {\bibinfo {author} {\bibfnamefont {S.}~\bibnamefont
  {{Nojiri}}}\ and\ \bibinfo {author} {\bibfnamefont {S.~D.}\ \bibnamefont
  {{Odintsov}}},\ }\href {\doibase 10.1016/j.physletb.2018.01.078} {\bibfield
  {journal} {\bibinfo  {journal} {Physics Letters B}\ }\textbf {\bibinfo
  {volume} {779}},\ \bibinfo {pages} {425} (\bibinfo {year} {2018})},\ \Eprint
  {http://arxiv.org/abs/1711.00492} {arXiv:1711.00492} \BibitemShut {NoStop}%
\bibitem [{\citenamefont {Avelino}(2012{\natexlab{b}})}]{nuclear.test}%
  \BibitemOpen
  \bibfield  {author} {\bibinfo {author} {\bibfnamefont {P.}~\bibnamefont
  {Avelino}},\ }\href@noop {} {\bibfield  {journal} {\bibinfo  {journal}
  {Journal of Cosmology and Astroparticle Physics}\ }\textbf {\bibinfo {volume}
  {2012}},\ \bibinfo {pages} {022} (\bibinfo {year}
  {2012}{\natexlab{b}})}\BibitemShut {NoStop}%
\bibitem [{\citenamefont {{Beltr{\'a}n Jim{\'e}nez}}\ \emph
  {et~al.}(2017)\citenamefont {{Beltr{\'a}n Jim{\'e}nez}}, \citenamefont
  {{Heisenberg}}, \citenamefont {{Olmo}},\ and\ \citenamefont
  {{Rubiera-Garcia}}}]{jimenez2017oct}%
  \BibitemOpen
  \bibfield  {author} {\bibinfo {author} {\bibfnamefont {J.}~\bibnamefont
  {{Beltr{\'a}n Jim{\'e}nez}}}, \bibinfo {author} {\bibfnamefont
  {L.}~\bibnamefont {{Heisenberg}}}, \bibinfo {author} {\bibfnamefont {G.~J.}\
  \bibnamefont {{Olmo}}}, \ and\ \bibinfo {author} {\bibfnamefont
  {D.}~\bibnamefont {{Rubiera-Garcia}}},\ }\href {\doibase
  10.1088/1475-7516/2017/10/029} {\bibfield  {journal} {\bibinfo  {journal}
  {JCAP}\ }\textbf {\bibinfo {volume} {10}},\ \bibinfo {eid} {029} (\bibinfo
  {year} {2017})},\ \Eprint {http://arxiv.org/abs/1707.08953} {arXiv:1707.08953
  [hep-th]} \BibitemShut {NoStop}%
\end{thebibliography}%
\end{document}